\documentclass[twocolumn,showpacs,prd]{revtex4}
\usepackage{mathrsfs}
\usepackage{longtable,lscape}\usepackage{multirow}
\usepackage{txfonts}
\usepackage{amssymb}
\usepackage{indentfirst}
\usepackage{graphicx,,booktabs}
\usepackage{color}
\usepackage{amssymb}

\begin{document}

\title{$Z_b(10610)^\pm$ and $Z_b(10650)^\pm$ as the $B^*\bar{B}$ and $B^*\bar{B}^{*}$ molecular states}
\author{Zhi-Feng Sun$^{1,2}$}
\author{Jun He$^{1,3}$}
\author{Xiang Liu$^{1,2}$\footnote{Corresponding author}}\email{xiangliu@lzu.edu.cn}
\affiliation{ $^1$Research Center for Hadron and CSR Physics,
Lanzhou University and Institute of Modern Physics of CAS, Lanzhou 730000, China\\
$^2$School of Physical Science and Technology, Lanzhou University, Lanzhou 730000,  China\\
$^3$Nuclear Theory Group, Institute of Modern Physics of CAS,
Lanzhou 730000, China}
\author{Zhi-Gang Luo}\author{Shi-Lin Zhu\footnote{Corresponding author}}
\email{zhusl@pku.edu.cn} \affiliation{Department of Physics
and State Key Laboratory of Nuclear Physics and Technology\\
Peking University, Beijing 100871, China}

\date{\today}

\begin{abstract}

In the framework of the one-boson-exchange model, we have studied
the interaction of the $B^*\bar{B}$ and $B^*\bar{B}^{*}$ system.
After considering the S-wave and D-wave mixing, we notice that
both $Z_b(10610)^\pm$ and $Z_b(10650)^\pm$ can be interpreted as
the $B^*\bar{B}$ and $B^*\bar{B}^{*}$ molecular states quite
naturally. Within the same framework, there also exist several molecular charmonia including $X(3872)$
and several other molecular bottomonia, which are the partners of $Z_b(10610)$ and $Z_b(10650)$.
The long-range one-pion-exchange force alone is strong enough to form these loosely bound
molecular states, which ensures the numerical results quite model-independent and robust.
\end{abstract}

\pacs{14.40.Rt, 12.39.Jh, 12.39.Hg, 14.40.Nd, 14.40.Lb} \maketitle

\section{Introduction}\label{sec1}

Very recently, the Belle Collaboration announced two charged
bottomonium-like states $Z_b(10610)$ and $Z_b(10650)$. These two
states were observed in the invariant mass spectra of
$h_b(nP)\pi^\pm$ ($n=1,2$) and $\Upsilon(mS)\pi^\pm$ ($m=1,2,3$)
of the corresponding $\Upsilon(5S)\to h_b(nP)\pi^+\pi^-$ and
$\Upsilon(5S)\to \Upsilon(mS)\pi^+\pi^-$ hidden-bottom decays
\cite{Collaboration:2011gj}. With the above five hidden-bottom
decay channels, Belle extracted the $Z_b(10610)$ and $Z_b(10650)$
parameters. The obtained averages over all five channels are
$M_{Z_b(10610)}=10608.4\pm2.0$ MeV/c$^2$,
$\Gamma_{Z_b(10610)}=15.6\pm2.5$ MeV/c$^2$,
$M_{Z_b(10650)}=10653.2\pm1.5$ MeV/c$^2$,
$\Gamma_{Z_b(10650)}=14.4\pm3.2$ MeV/c$^2$
\cite{Collaboration:2011gj}. In addition, the analysis of the
angular distribution indicates both $Z_b(10610)$ and $Z_b(10650)$
favor $I^G(J^P)=1^+(1^+)$.

If $Z_b(10610)$ and $Z_b(10650)$ arise from the resonance
structures, they are good candidates of non-conventional
bottomonium-like states. The masses of the $J^{PC}=1^{++}$ and
$J^{PC}=1^{+-}$ $b\bar b q\bar q$ tetraquark states were found to
be around $10.1\sim 10.2$ GeV in the framework of QCD sum rule
formalism \cite{chenwei}, which are significantly lower than these
two charged $Z_b$ states. Therefore, it's hard to accommodate them
as tetraquarks. If comparing the experimental measurement with the
$B\bar{B}^*$ and $B^*\bar{B}^{*}$ thresholds, one notices that
$Z_b(10610)$ and $Z_b(10650)$ are close to thresholds of
$B\bar{B}^*$ and $B^*\bar{B}^{*}$, respectively. One plausible
explanation is that both $Z_b(10610)$ and $Z_b(10650)$ are either
$B^*\bar{B}^{*}$ or $B^*\bar{B}^{*}$ molecular states
respectively.

Before the observations of two charged $Z_b(10610)$ and
$Z_b(10650)$ states, there have been many theoretical works which
focused on the molecular systems composed of $B^{(*)}$ and
$\bar{B}^{(*)}$ meson pair and indicated that there probably exist
loosely bound S-wave $B^*\bar{B}^{*}$ or $B^*\bar{B}^{*}$
molecular states \cite{Liu:2008fh,Liu:2008tn}. To some extent,
such studies were stimulated by a series of near-threshold
charomonium-like $X$, $Y$, $Z$ states in the past eight years.

Molecular states involving charmed quarks were first proposed by
Voloshin and Okun more than thirty years ago
\cite{Voloshin:1976ap}. Later, De Rujula, Georgi and Glashow
speculated $\psi(4040)$ as a $D^*\bar{D}^*$ molecular charmonium
\cite{De Rujula:1976qd}. T\"{o}nqvist calculated the possible
deuteron-like two-meson bound states such as $D\bar{D}^*$ and
$D^*\bar{D}^*$ using the quark-pion interaction model
\cite{Tornqvist:1993vu,Tornqvist:1993ng}. The observations of
$X(3872)$, three charged charomonium-like states $Z^+(4350)$,
$Z_1^+(4050)$, $Z^+(4250)$ and $Y(4140)$, $Y(4274)$ etc. again
inspired theorists' interest in the molecular system composed of
charmed meson pair (see Refs.
\cite{Close:2003sg,Voloshin:2006pz,Wong:2003xk,Swanson:2003tb,Swanson:2004pp,Tornqvist:2004qy,
Suzuki:2005ha,Liu:2008fh,Liu:2008tn,Liu:2008du,Liu:2008qb,Liu:2008mi,Liu:2009zzf,Liu:2010xh,
Thomas:2008ja,Lee:2009hy,Meng:2007fu,Liu:2007bf,Liu:2008xz,Ding:2007ar,Ding:2008mp,Ding:2008gr,
Ding:2009vj,Ding:2009vd,Ding:2009zq,Liu:2007ez,Shen:2010ky,Hu:2010fg,Liu:2009ei,Liu:2010hf,Yang:2011wz}
for details).

As the first observed charged bottomonium-like states,
$Z_b(10610)$ and $Z_b(10650)$ have attracted the attention of many
theoretical groups. The authors discussed the special decay
behavior of the $J=1$ S-wave $B\bar{B}^*$ and $B^*\bar{B}^*$
molecular states based on the heavy quark symmetry in Ref.
\cite{Bondar:2011ev}. Chen, Liu and Zhu \cite{Chen:2011zv} found
that the intermediate $Z_b(10610)$ and $Z_b(10650)$ contribution
to $\Upsilon(5S)\to \Upsilon(2S)\pi^+\pi^-$ naturally explains
Belle's previous observation of the anomalous
$\Upsilon(2S)\pi^+\pi^-$ production near the peak of
$\Upsilon(5S)$ at $\sqrt s=10.87$ GeV \cite{Abe:2007tk}, where the
resulting $d\Gamma(\Upsilon(5S)\to
\Upsilon(2S)\pi^+\pi^-)/dm_{\pi^+\pi^-}$ and
$d\Gamma(\Upsilon(5S)\to \Upsilon(2S)\pi^+\pi^-)/d\cos\theta$
distributions agree with Belle's measurement after inclusion of
these $Z_b$ states \cite{Chen:2011zv}. The authors of Ref.
\cite{Zhang:2011jj} tried to reproduce the masses of $Z_b(10610)$
and $Z_b(10650)$ using a molecular bottomonium-like current in the
QCD sum rule calculation. Yang {\it et al.} studied the mass
spectra of the S-wave $[\bar{b}q][b\bar{q}]$,
$[\bar{b}q]^*[b\bar{q}]$, $[\bar{b}q]^*[b\bar{q}]^*$ in the chiral
quark model and indicated that $Z_b(10610)$ and $Z_b(10650)$ are
good candidates of the S-wave $B\bar{B}^*$ and $B^*\bar{B}^*$
bound states \cite{Yang:2011rp}. Bugg proposed a non-exotic
explanation of $Z_b(10610)$ and $Z_b(10650)$, which are
interpreted as the orthogonal linear combinations of the $q\bar q$
and meson-meson states, namely $b\bar{b} + B\bar{B}^*$ and
$b\bar{b} + B^*\bar{B}^*$ \cite{Bugg:2011jr}, respectively. Nieves
and Valderrama suggested the possible existence of two positive
C-parity isoscalar states: a $^3S_1-^3D_1$ state with a binding
energy of 90-100 MeV and a $^3P_0$ state located about 20-30 MeV
below the $B\bar{B}^*$ threshold \cite{nieves}. Unfortunately, the
quantum number of the above states does not match those of these
two charged $Z_b$ states. Danilkin, Orlovsky and Simonov studied
the interaction between a light hadron and heavy quarkonium
through the transition to a pair of intermediate heavy mesons.
Based on the above coupled-channel effect, the authors discussed
the resonance structures close to the $B^{(*)}\bar B^\ast$
threshold \cite{russian}. Using the chromomagnetic interaction,
the authors of Ref. \cite{Guo:2011gu} discussed the possibility of
$Z_b(10610)$ and $Z_b(10650)$ being tetraquark states. In
contrast, the $b\bar b q\bar q$ tetraquark states were predicted
to be around $10.2 \sim 10.3$ GeV using the color-magnetic
interaction with the flavor symmetry breaking corrections
\cite{cui}, consistent with the values extracted from the QCD sum
rule approach \cite{chenwei}.

As emphasized in Ref. \cite{Chen:2011zv}, future dynamical study
of the mass and decay pattern of the S-wave $B\bar{B}^*$ and
$B^*\bar{B}^*$ molecular states are very desirable. In this work,
we perform more thorough study of the $B\bar{B}^*$ and
$B^*\bar{B}^*$ systems using the One-Boson-Exchange (OBE) model.
Different from our former work in Refs.
\cite{Liu:2008fh,Liu:2008tn}, we not only consider S-wave
interaction but also include D-wave contribution between $B^{(*)}$
and $\bar{B}^{(*)}$. Such a study will be helpful to answer
whether the $B\bar{B}^*$ and $B^*\bar{B}^*$ molecular bottomonia
exist or not.

This paper is organized as follows. After the introduction, we
present the formalism of the study of the $B\bar{B}^*$ and
$B^*\bar{B}^*$ systems, which includes the relevant effective
Lagrangian and coupling constants, the derivation of the effective
potential of the $B\bar{B}^*$ and $B^*\bar{B}^*$ system, the
numerical results etc.. Finally, the paper ends with the
discussion and conclusion.

\section{Deduction of effective potential}\label{sec2}

\subsection{Flavor wave function}

We list the flavor wave functions of the $B\bar{B}^*$ and
$B^*\bar{B}^*$ systems constructed in Refs.
\cite{Liu:2008fh,Liu:2008tn}. The $B\bar{B}^*$ systems can be
categorized as the isovector and isoscalar states with the
corresponding flavor wave functions
\begin{eqnarray}
&&\left\{\begin{array}{l}
|{Z_{B\bar{B}^*}^{(T)}}^+\rangle=\frac{1}{\sqrt{2}}\big(|B^{*+}\bar{B}^0\rangle+cB^+\bar{B}^{*0}\big),\\
|{Z_{B\bar{B}^*}^{(T)}}^-\rangle=\frac{1}{\sqrt{2}}\big(|B^{*-}\bar{B}^0\rangle+cB^-\bar{B}^{*0}\big),\\
|{Z_{B\bar{B}^*}^{(T)}}^0\rangle=\frac{1}{2}\Big[\big(|B^{*+}B^-\rangle-B^{*0}\bar{B}^0\big)
+c\big(B^+B^{*-}-B^0\bar{B}^{*0}\big)\Big],\end{array}\right. \label{e1}\\
&&|{Z_{B\bar{B}^*}^{(S)}}^0\rangle=\frac{1}{2}\Big[\big(|B^{*+}B^-\rangle+B^{*0}\bar{B}^0\big)
+c\big(B^+B^{*-}+B^0\bar{B}^{*0}\big)\Big],\label{e2}
\end{eqnarray}
where $c=\pm$ corresponds to C-parity $C=\mp$ respectively
\cite{Liu:2008fh,Liu:2008tn}. The flavor wave functions of the
$B^*\bar{B}^*$ systems can be constructed as
\begin{eqnarray}
\left\{\begin{array}{l}
|{Z_{B^*\bar{B}^*}^{(T)}}[\mathrm{J}]^+\rangle=|B^{*+}\bar{B}^{*0}\rangle\\
|{Z_{B^*\bar{B}^*}^{(T)}}[\mathrm{J}]^-\rangle=|B^{*-}\bar{B}^{*0}\rangle\\
|{Z_{B^*\bar{B}^*}^{(T)}}[\mathrm{J}]^0\rangle=\frac{1}{\sqrt{2}}\big(|B^{*+}B^{*-}\rangle-|B^{*0}\bar{B}^{*0}\rangle\big)
\end{array}\right.\label{e3}
\end{eqnarray}
for the isovector states, and
\begin{eqnarray}
|{Z_{B^*\bar{B}^*}^{(S)}}[\mathrm{J}]^0\rangle&=&\frac{1}{\sqrt{2}}\big(|B^{*+}B^{*-}\rangle+|B^{*0}\bar{B}^{*0}\rangle\big)\label{e4}
\end{eqnarray}
for the isoscalar state. In the above expressions, the
superscripts $T$ and $S$ in Eqs. (\ref{e1})-(\ref{e4}) are applied
to distinguish the isovector and isoscalar states, respectively.
The total angular momentum of the S-wave $B^*\bar{B}^*$ systems is
$J=0,1,2$. Thus, we use the extra notation $[\mathrm{J}]$ in Eqs.
(\ref{e3})-(\ref{e4}) to distinguish the $B^*\bar{B}^*$ systems
with different total angular momentum $J$.

Belle indicated that both $Z_b(10610)$ and $Z_b(10650)$ belong to
the isotriplet states. If $Z_b(10610)$ and $Z_b(10650)$ are the
$B\bar{B}^*$ or $B^*\bar{B}^*$ molecular states respectively, they
should correspond to ${Z_{B\bar{B}^*}^{(T)}}$ and
${Z_{B^*\bar{B}^*}^{(T)}}[1]$ in Eqs. (\ref{e1}) and (\ref{e3}),
respectively. Since $Z_b(10610)^0$ is of $C$-odd parity, i.e.,
$C=-1$, thus the coefficient $c=+1$ is taken in Eq. (\ref{e1}).
The choice of the coefficient $c=-1$ and $C=+1$ leads to $X(3872)$
and its partners, where $X(3872)$ corresponds to
${Z_{D\bar{D}^*}^{(S)}}^\prime$ listed in Table. \ref{wave}.

In Table \ref{wave}, we summarize the quantum numbers of the
states when we discuss whether there exist the $B\bar{B}^*$ and
$B^*\bar{B}^*$ molecular states. Moreover, we extend the same
formalism to study the $D\bar{D}^*$ and $D^*\bar{D}^*$ systems,
where the flavor wave function of the $D\bar{D}^*$ and
$D^*\bar{D}^*$ systems can be obtained with replacement
$B^{(*)}\to \bar{D}^{(*)}$ and $\bar{B}^{(*)}\to {D}^{(*)}$.

\renewcommand{\arraystretch}{1.6}
\begin{table}[hbt]
\caption{A summary of the $B\bar{B}^*$, $B^*\bar{B}^*$,
$D\bar{D}^*$, $D^*\bar{D}^*$ systems. If taking $c=-1$ in Eqs.
(\ref{e1}) and (\ref{e2}), we obtain the flavor wave functions of
${Z_{B\bar{B}^*}^{(T)}}^\prime$ and
${Z_{B\bar{B}^*}^{(S)}}^\prime$, which are the partners of
${Z_{B\bar{B}^*}^{(T)}}$ and ${Z_{B\bar{B}^*}^{(S)}}$
respectively. \label{wave}}
\begin{tabular}{ccc}\toprule[1pt]
$B\bar{B}^*/B^*\bar{B}^*$ systems& $D\bar{D}^*/D^*\bar{D}^*$
systems&$I^G(J^{PC})$\\\midrule[1pt]
$Z_{B\bar{B}^*}^{(T)}$& $Z_{D\bar{D}^*}^{(T)}$& $1^+(1^+)$\\
$Z_{B\bar{B}^*}^{(S)}$&$Z_{D\bar{D}^*}^{(S)}$&$0^-(1^{+-})$\\
$Z_{B^*\bar{B}^*}^{(T)}[\mathrm{J}]$&$Z_{D^*\bar{D}^*}^{(T)}[\mathrm{J}]$&$1^-(\mathrm{0}^+),1^-(\mathrm{2}^+),1^+(\mathrm{1}^+)$\\
$Z_{B^*\bar{B}^*}^{(S)}[\mathrm{J}]$&$Z_{D^*\bar{D}^*}^{(S)}[\mathrm{J}]$&$0^+(\mathrm{0}^{++}),0^+(\mathrm{2}^{++}),0^-(\mathrm{1}^{+-})$\\
${Z_{B\bar{B}^*}^{(T)}}^\prime$&${Z_{D\bar{D}^*}^{(T)}}^\prime$&$1^-(1^+)$\\
${Z_{B\bar{B}^*}^{(S)}}^\prime$&${Z_{D\bar{D}^*}^{(S)}}^\prime$&$0^+(1^{++})$\\
\bottomrule[1pt]
\end{tabular}
\end{table}

\subsection{Effective Lagrangian and coupling constant}

In order to obtain the effective potential of the $B\bar{B}^*$ and
$B^*\bar{B}^{*}$ system, we employ the OBE model, which is an
effective framework to describe the $B\bar{B}^*$ or
$B^*\bar{B}^{*}$ interaction by exchanging the light pseudoscalar,
scalar and vector mesons. In terms of heavy quark limit and chiral
symmetry, the interactions of light pesudoscalar, vector and
scalar mesons interacting with S-wave heavy flavor mesons were
constructed in Refs.
\cite{Cheng:1992xi,Yan:1992gz,Wise:1992hn,Burdman:1992gh,Casalbuoni:1996pg,Falk:1992cx,Ding:2008gr}

\begin{eqnarray}
\mathcal{L}_{HH\mathbb{P}}&=& ig\langle H^{(Q)}_b \gamma_\mu
A_{ba}^\mu\gamma_5 \bar{H}^{(Q)}_a\rangle
\nonumber\\
&&+ ig\langle
\bar{H}^{(\bar{Q})}_a \gamma_\mu A_{ab}^\mu\gamma_5H_b^{(\bar{Q})}\rangle,\label{eq:lag}\\
\mathcal{L}_{HH\mathbb{V}}&=& i\beta\langle H^{(Q)}_b v_\mu
(\mathcal{V}^\mu_{ba}-\rho^\mu_{ba})\bar{H}^{(Q)}_a\rangle\nonumber\\
&&+i\lambda\langle H^{(Q)}_b
\sigma_{\mu\nu}F^{\mu\nu}(\rho)\bar{H}^{(Q)}_a\rangle\nonumber\\
&&-i\beta\langle \bar{H}^{(\bar{Q})}_a v_\mu
(\mathcal{V}^\mu_{ab}-\rho^\mu_{ab})H^{(\bar{Q})}_b\rangle\nonumber\\
&&+i\lambda\langle H_b^{(\bar{Q})}
\sigma_{\mu\nu}F'^{\mu\nu}(\rho)\bar{H}^{(\bar{Q})}_a\rangle,\\
\mathcal{L}_{ HH\sigma}&=&g_s \langle H^{(Q)}_a\sigma
\bar{H}^{(Q)}_a\rangle+g_s \langle \bar{H}^{(\bar{Q})}_a\sigma
H^{(\bar{Q})}_a\rangle, \label{eq:lag2}
\end{eqnarray}
where the multiplet field $H^{(Q)}$ is composed of the
pseudoscalar ${\mathcal{P}}$ and vector ${\mathcal{P}}^{*}$ with
${\mathcal{P}}^{(*)T} =(D^{(*)+},D^{(*)0})$ or
$(\bar{B}^{(*)0},B^{(*)-})$. And $H^{(Q)}$ and $\bar H^{(Q)}$ are
defined by
\begin{eqnarray}
    H_a^{(Q)}&=&\frac{1+\rlap\slash
    v}{2}[{\mathcal{P}}^{*}_{a\mu}\gamma^\mu
    -{\mathcal{P}}_a\gamma_5],\\
    \bar{H}_a^{(Q)}&=&[{\mathcal{P}}^{*\dag}_{a\mu}\gamma^\mu
    +{\mathcal{P}}_a^{\dag}\gamma_5]\frac{1+\rlap\slash
    v}{2}.
\end{eqnarray}
Here, $\bar{H}=\gamma_0H^\dag\gamma_0$ and $v=(1,{\mathbf 0})$.

As given in Refs.~\cite{Ding:2008gr,Grinstein:1992qt}, the
anti-charmed or bottom meson fields
$\widetilde{\mathcal{P}}^{(*)T}=(D^{(*)-},\bar{D}^{(*)0})$ or
$(B^{(*)0},B^{(*)+})$ satisfy
\begin{eqnarray}
    &&\widetilde{P}^{*}_{a\mu}=-\mathcal{C}{P}^{*}_{a\mu}\mathcal{C}^{-1},\
    \widetilde{P}_{a}=\mathcal{C}{P}_{a}\mathcal{C}^{-1}.
\end{eqnarray}
The multiplet field $H^{(\bar{Q})}$ with the heavy antiquark can
be defined as
\begin{eqnarray}
    && H_a^{(\bar{Q})}=C(\mathcal{C}H_a^{(Q)}
    \mathcal{C}^{-1})^TC^{-1}
    =[\widetilde{P}_a^{*\mu}\gamma_\mu-\widetilde{P}_a\gamma_5]\
    \frac{1-\rlap\slash v}{2},\\
    &&\bar{H}_a^{(\bar{Q})}
    =\frac{1-\rlap\slash
    v}{2}[\widetilde{P}_a^{*\mu}\gamma_\mu
    +\widetilde{P}_a\gamma_5].
\end{eqnarray}

If considering the following charge conjugation transformation,
\begin{eqnarray}
    &&\mathcal{C}\xi\mathcal{C}^{-1}=\xi^T,\ \
    \mathcal{C}\mathcal{V}_\mu
    \mathcal{C}^{-1}=-\mathcal{V}^T_\mu,\nonumber\\
    &&\mathcal{C}\mathcal{A}_\mu\mathcal{C}^{-1}=\mathcal{A}_\mu^T, \
    \ \
    \mathcal{C}\rho^\mu\mathcal{C}^{-1}=-\rho^{\mu T},
\end{eqnarray}
one obtains the Lagrangian relevant to the mesons with heavy
antiquark $\bar{Q}$ which is converted from the one related to the
meson with heavy quark $Q$, where the Lagrangians are given in
Eqs.
(\ref{eq:lag})-(\ref{eq:lag2})~\cite{Ding:2008gr,Grinstein:1992qt}.
In the above expressions, the
$\mathcal{P}(\widetilde{\mathcal{P}})$ and
$\mathcal{P}^*(\widetilde{\mathcal{P}}^*)$ satisfy the
normalization relations $\langle
0|{\mathcal{P}}|Q\bar{q}(0^-)\rangle =\langle
0|\widetilde{\mathcal{P}}|\bar{Q}q(0^-)\rangle
=\sqrt{M_\mathcal{P}}$ and $\langle
0|{\mathcal{P}}^*_\mu|Q\bar{q}(1^-)\rangle=\langle
0|\widetilde{\mathcal{P}}^{*}_\mu|\bar{Q}q(1^-)\rangle=\epsilon_\mu\sqrt{M_{\mathcal{P}^*}}$.
The axial current is
$A^\mu=\frac{1}{2}(\xi^\dag\partial_\mu\xi-\xi \partial_\mu
\xi^\dag)=\frac{i}{f_\pi}\partial_\mu{\mathbb P}+\cdots$ with
$\xi=\exp(i\mathbb{P}/f_\pi)$ and $f_\pi=132$ MeV.
$\rho^\mu_{ba}=ig_{V}\mathbb{V}^\mu_{ba}/\sqrt{2}$,
$F_{\mu\nu}(\rho)=\partial_\mu\rho_\nu - \partial_\nu\rho_\mu +
[\rho_\mu,{\ } \rho_\nu]$, $F'_{\mu\nu}(\rho)=\partial_\mu\rho_\nu
-
\partial_\nu\rho_\mu -
[\rho_\mu,{\ } \rho_\nu]$ and $g_V=m_\rho/f_\pi$, with $g_V=5.8$.
Here, $\mathbb P$ and $\mathbb V$ are two by two pseudoscalar and
vector matrices
\begin{eqnarray}
    {\mathbb P}&=&\left(\begin{array}{ccc}
        \frac{1}{\sqrt{2}}\pi^0+\frac{\eta}{\sqrt{6}}&\pi^+\\
        \pi^-&-\frac{1}{\sqrt{2}}\pi^0+\frac{\eta}{\sqrt{6}}
\end{array}\right), \\
\mathbb{V}&=&\left(\begin{array}{ccc}
\frac{\rho^{0}}{\sqrt{2}}+\frac{\omega}{\sqrt{2}}&\rho^{+}\\
\rho^{-}&-\frac{\rho^{0}}{\sqrt{2}}+\frac{\omega}{\sqrt{2}}
\end{array}\right).
\end{eqnarray}

By expanding Eqs. (\ref{eq:lag})-(\ref{eq:lag2}), one further
obtains the effective Lagrangian of the light pseudoscalar mesons
$\mathbb{P}$ with the heavy flavor mesons
\begin{eqnarray}\label{eq:lag-p-exch}
\mathcal{L}_{\mathcal{P}^*\mathcal{P}^*\mathbb{P}} &=&
-i\frac{2g}{f_\pi}\varepsilon_{\alpha\mu\nu\lambda}
v^\alpha\mathcal{P}^{*\mu}_{b}{\mathcal{P}}^{*\lambda\dag}_{a}
\partial^\nu{}\mathbb{P}_{ba}\nonumber\\
&&+i \frac{2g}{f_\pi}\varepsilon_{\alpha\mu\nu\lambda}
v^\alpha\widetilde{\mathcal{P}}^{*\mu\dag}_{a}\widetilde{\mathcal{P}}^{*\lambda}_{b}
\partial^\nu{}\mathbb{P}_{ab},\label{ppp}\\
\mathcal{L}_{\mathcal{P}^*\mathcal{P}\mathbb{P}} &=&-
\frac{2g}{f_\pi}(\mathcal{P}^{}_b\mathcal{P}^{*\dag}_{a\lambda}+
\mathcal{P}^{*}_{b\lambda}\mathcal{P}^{\dag}_{a})\partial^\lambda{}
\mathbb{P}_{ba}\nonumber\\
&&+\frac{2g}{f_\pi}(\widetilde{\mathcal{P}}^{*\dag}_{a\lambda}\widetilde{\mathcal{P}}_b+
\widetilde{\mathcal{P}}^{\dag}_{a}\widetilde{\mathcal{P}}^{*}_{b\lambda})\partial^\lambda{}\mathbb{P}_{ab}.
\end{eqnarray}

The effective Lagrangian depicting the coupling of the light
vector mesons $\mathbb{V}$ and heavy flavor mesons reads as
\begin{eqnarray}\label{eq:lag-v-exch}
  \mathcal{L}_{\mathcal{PP}\mathbb{V}}
  &=& -\sqrt{2}\beta{}g_V\mathcal{P}^{}_b\mathcal{P}_a^{\dag}
  v\cdot\mathbb{V}_{ba}
 +\sqrt{2}\beta{}g_V\widetilde{\mathcal{P}}^{\dag}_a
  \widetilde{\mathcal{P}}^{}_b
  v\cdot\mathbb{V}_{ab},\label{ppv}\nonumber\\\\
  \mathcal{L}_{\mathcal{P}^*\mathcal{P}\mathbb{V}}
  &=&- 2\sqrt{2}\lambda{}g_V v^\lambda\varepsilon_{\lambda\mu\alpha\beta}
  (\mathcal{P}^{}_b\mathcal{P}^{*\mu\dag}_a +
  \mathcal{P}_b^{*\mu}\mathcal{P}^{\dag}_a)
  (\partial^\alpha{}\mathbb{V}^\beta)_{ba}\nonumber\\
&&-  2\sqrt{2}\lambda{}g_V
v^\lambda\varepsilon_{\lambda\mu\alpha\beta}
(\widetilde{\mathcal{P}}^{*\mu\dag}_a\widetilde{\mathcal{P}}^{}_b
+
\widetilde{\mathcal{P}}^{\dag}_a\widetilde{\mathcal{P}}_b^{*\mu})
  (\partial^\alpha{}\mathbb{V}^\beta)_{ab},\label{ppsv}\nonumber\\\\
  \mathcal{L}_{\mathcal{P}^*\mathcal{P}^*\mathbb{V}}
  &=& \sqrt{2}\beta{}g_V \mathcal{P}_b^{*}\cdot\mathcal{P}^{*\dag}_a
  v\cdot\mathbb{V}_{ba}\nonumber\\
  &&-i2\sqrt{2}\lambda{}g_V\mathcal{P}^{*\mu}_b\mathcal{P}^{*\nu\dag}_a
  (\partial_\mu{}
  \mathbb{V}_\nu - \partial_\nu{}\mathbb{V}_\mu)_{ba}\nonumber\\
  &&-\sqrt{2}\beta g_V
  \widetilde{\mathcal{P}}^{*\dag}_a\widetilde{\mathcal{P}}_b^{*}
  v\cdot\mathbb{V}_{ab}\nonumber\\
  &&-i2\sqrt{2}\lambda{}g_V\widetilde{\mathcal{P}}^{*\mu\dag}_a\widetilde{\mathcal{P}}^{*\nu}_b(\partial_\mu{}
  \mathbb{V}_\nu - \partial_\nu{}\mathbb{V}_\mu)_{ab}.\label{pspsv}
  \end{eqnarray}

The effective Lagrangian of the scalar meson $\sigma$ interacting
with the heavy flavor mesons can be expressed as
\begin{eqnarray}\label{eq:lag-s-exch}
  \mathcal{L}_{\mathcal{PP}\sigma}
  &=& -2g_s\mathcal{P}^{}_b\mathcal{P}^{\dag}_b\sigma
 -2g_s\widetilde{\mathcal{P}}^{}_b\widetilde{\mathcal{P}}^{\dag}_b\sigma,\\
  \mathcal{L}_{\mathcal{P}^*\mathcal{P}^*\sigma}
  &=& 2g_s\mathcal{P}^{*}_b\cdot{}\mathcal{P}^{*\dag}_b\sigma
 +2g_s\widetilde{\mathcal{P}}^{*}_b\cdot{}\widetilde{\mathcal{P}}^{*\dag}_b\sigma.
\end{eqnarray}

As shown in Eqs. (\ref{ppp})-(\ref{pspsv}), the terms for the
interactions between the anti-heavy flavor mesons and light mesons
can be obtained by taking the following replacements in the
corresponding terms for the interactions between the heavy flavor
mesons and light mesons:
\begin{eqnarray}
  v\to -v,\; a\to b,\; b\to a, \nonumber \\
  \mathcal{P}^*_{\mu}\to \widetilde{\mathcal{P}}^{*\dag}_\mu,\; \mathcal{P}\to -\widetilde{\mathcal{P}}^\dag,\nonumber\\
  \mathcal{P}^{*\dag}_\mu\to \widetilde{\mathcal{P}}^*_\mu,\; \mathcal{P}^\dag\to - \widetilde{\mathcal{P}}.\nonumber
\end{eqnarray}

$g=0.59$ is extracted from the experimental width of $D^{*+}$
~\cite{Isola:2003fh}. The parameter $\beta$ relevant to the vector
meson can be fixed as $\beta=0.9$ by the vector meson dominance
mechanism while $\lambda=0.56$ GeV$^{-1}$ was obtained by
comparing the form factor calculated by light cone sum rule with
the one obtained by lattice QCD. As the coupling constant related
to the scalar meson $\sigma$, $g_s=g_\pi/(2\sqrt{6})$ with
$g_\pi=3.73$ was given in Refs. \cite{Falk:1992cx,Liu:2008tn}.

\subsection{Effective potential}

With the above preparation, we deduce the effective potentials of
the $B\bar{B}^*$ and $B^*\bar{B}^*$ systems in the following .
Generally, the scattering amplitude $i\mathcal{M}(J,J_Z)$ is
related to the interaction potential in the momentum space in
terms of the Breit approximation
\begin{eqnarray*}
\mathcal{V}_E^{B^{(*)}\bar{B}^{(*)}}(\mathbf{q})&=&-\frac{\mathcal{M}({B^{(*)}\bar{B}^{(*)}}\to
{B^{(*)}\bar{B}^{(*)}})}{\sqrt{\prod_i 2M_i \prod_f 2M_f}},
\end{eqnarray*}
where $M_{i}$ and $M_j$ denote the masses of the initial and final
states respectively. The potential in the coordinate space
$\mathcal{V}(\mathbf{r})$ is obtained after performing the Fourier
transformation
\begin{eqnarray}
\mathcal{V}_E^{B^{(*)}\bar{B}^{(*)}}(\mathbf{r})=\int\frac{d\mathbf{p}}{(2\pi)^3}\,e^{i
\mathbf{p}\cdot
\mathbf{r}}\mathcal{V}_E^{B^{(*)}\bar{B}^{(*)}}(\mathbf{q})\mathcal{F}^2(q^2,m_E^2),
\end{eqnarray}
where we need to introduce the monopole form factor (FF)
$\mathcal{F}(q^2,m_E^2)=({\Lambda^2-m_E^2})/({\Lambda^2-q^2})$ to
reflect the structure effect of the vertex of the heavy mesons
interacting with the light mesons. $m_E$ denotes the exchange
meson mass. For $q^2\to 0$ we can treat FF as a constant while for
$\Lambda\gg m$ FF approaches unity. The behavior of FF indicates
\cite{Liu:2008fh} (1) when the distance becomes infinitely large,
the interaction vertex looks like a perfect point corresponding of
the constant FF; (2) when the distance is very small, the inner
structure would manifest itself. In reality, the phenomenological
cutoff $\Lambda$ is around one to several GeV, which also plays
the role of regulating the effective potential.

In this work, we consider both S-wave and D-wave interactions
between $B^{(*)}$ and $\bar{B}^{(*)}$ mesons. In general, the
$B\bar{B}^{*}$ and $B^*\bar{B}^{*}$ states can be expressed as
\begin{eqnarray}
\big|{Z_{B\bar{B}^*}^{(\alpha)}}^{(\prime)}\big\rangle&=&\left
(\begin{array}{c}
\Big|BB^*(^3S_1)\Big\rangle\\
\Big|BB^*(^3D_1)\Big\rangle\end{array}\right ),\,
\big|Z_{B^*\bar{B}^*}^{(\alpha)}[\mathrm{0}]\big\rangle=\left
(\begin{array}{c}
\Big|B^*\bar{B}^*(^1S_0)\Big\rangle\\
\Big|B^*\bar{B}^*(^5D_0)\Big\rangle,
\end{array}\right )\nonumber\\
\big|Z_{B^*\bar{B}^*}^{(\alpha)}[\mathrm{1}]\big\rangle&=&\left
(\begin{array}{c}
\Big|B^*\bar{B}^*(^3S_1)\Big\rangle\\
\Big|B^*\bar{B}^*(^3D_1)\Big\rangle\\
\Big|B^*\bar{B}^*(^5D_1)\Big\rangle\end{array}\right ),\,
\big|Z_{B^*\bar{B}^*}^{(\alpha)}[\mathrm{2}]\big\rangle=\left
(\begin{array}{c}
\Big|B^*\bar{B}^*(^5S_2)\Big\rangle\\
\Big|B^*\bar{B}^*(^1D_2)\Big\rangle\\
\Big|B^*\bar{B}^*(^3D_2)\Big\rangle\\
\Big|B^*\bar{B}^*(^5D_2)\Big\rangle\end{array}\right )\nonumber\\
\label{11}
\end{eqnarray}
with $\alpha=S,T$, where we use the notation $^{2S+1}L_J$ to
denote the total spin $S$, angular momentum $L$, total angular
momentum $J$ of the $B\bar{B}^*$ or $B^*\bar{B}^*$ system. Indices
$S$ and $D$ indicate that the couplings between $B^*$ and
$\bar{B}^*$ occur via the $S$-wave and $D$-wave interactions,
respectively.

Thus, the total effective potentials of the $B\bar{B}^*$ and
$B^*\bar{B}^*$ systems are
\begin{eqnarray}
V_{\mathrm{Total}}^{{Z_{B\bar{B}^*}^{(\alpha)}}^{(\prime)}}&=&\Big\langle {Z_{B\bar{B}^*}^{(\alpha)}}^{(\prime)}\big)\Big| \sum_{E=\mathrm{\pi,\eta,\sigma,\rho,\omega}}\mathcal{V}_E^{B\bar{B}^*}(r)\big|{Z_{B\bar{B}^*}^{(\alpha)}}^{(\prime)}\big\rangle,\\
V_{\mathrm{Total}}^{{Z_{B^*\bar{B}^*}^{(\alpha)}}[\mathrm{J}]}&=&\Big\langle
{Z_{B^*\bar{B}^*}^{(\alpha)}}[\mathrm{J}]\big)\Big|
\sum_{E=\mathrm{\pi,\eta,\sigma,\rho,\omega}}\mathcal{V}_E^{B^*\bar{B}^*}(r)\big|{Z_{B^*\bar{B}^*}^{(\alpha)}}[\mathrm{J}]\big\rangle,
\end{eqnarray}
which are $2\times 2$ and $(J+2)\times (J+2)$ matrices
respectively. We impose the following constraint
\begin{eqnarray}
    \Big|B\bar{B}^*\big(^{2S+1}L_J\big)\Big\rangle&=&\sum_{m,m_L,m_S}C_{1m,Lm_L}^{JM}
    \epsilon^{m}_{n}Y_{Lm_L},\label{po0}\\
    \Big|B^*\bar{B}^*\big(^{2S+1}L_J\big)\Big\rangle&=&\sum_{m,m',m_L,m_S}C_{Sm_S,Lm_L}^{JM}
    C_{1m,1m'}^{Sm_S}\epsilon^{m'}_{n'}\epsilon^{m}_{n}Y_{Lm_L},\nonumber\\\label{po}
\end{eqnarray}
to the effective potential obtained from the scattering amplitude.
$C_{1m,Lm_L}^{JM}$, $C_{Sm_S,Lm_L}^{JM}$ and $C_{1m,1m'}^{Sm_S}$
are the Clebsch-Gordan coefficients. $Y_{Lm_L}$ is the spherical
harmonics function. The polarization vector for the vector heavy
flavor meson is defined as
$\epsilon^m_\pm=\mp\frac{1}{\sqrt{2}}(\epsilon^m_x\pm
i\epsilon^m_y)$ and $\epsilon^m_0=\epsilon^m_z$. Here, the
polarization vector in Eqs. (\ref{po0})-(\ref{po}) is just the one
appearing in the effective potentials which will be presented
later.

\subsubsection{The $B\bar{B}^*$ system}\label{bbs}

The general expressions of the total effective potentials of the
isoscalar and isovector $B\bar{B}^*$ systems are
\begin{eqnarray}
\mathcal{V}^{{Z_{B\bar{B}^*}^{(T)}}^{(\prime)}}&=&
V^{\mathrm{Direct}}_\sigma-\frac{1}{2}V^{\mathrm{Direct}}_\rho
+\frac{1}{2}V^{\mathrm{Direct}}_\omega+\frac{c}{4}\bigg(-2V^{\mathrm{Cross}}_\pi\nonumber\\&&+
\frac{2}{3}V^{\mathrm{Cross}}_
\eta-2V^{\mathrm{Corss}}_\rho+2 V^{\mathrm{Cross}}_\omega\bigg),\label{h1}\\
\mathcal{V}^{{Z_{B\bar{B}^*}^{(S)}}^{(\prime)}}&=&
V^{\mathrm{Direct}}_\sigma+\frac{3}{2}V^{\mathrm{Direct}}_\rho
+\frac{1}{2}V^{\mathrm{Direct}}_\omega+\frac{c}{4}\bigg(6V^{\mathrm{Cross}}_\pi\nonumber\\&&+
\frac{2}{3}V^{\mathrm{Cross}}_ \eta+6V^{\mathrm{Corss}}_\rho+2
V^{\mathrm{Corss}}_\omega\bigg),\label{h2}
\end{eqnarray}
where the subpotentials from the $\pi$, $\eta$, $\sigma$, $\rho$
and $\omega$ meson exchanges are written as
\begin{eqnarray}
V^{\mathrm{Cross}}_\pi&=&
-\frac{g^2}{f_\pi^2}\bigg[\frac{1}{3}(\mbox{\boldmath$\epsilon$}_2\cdot
\mbox{\boldmath$\epsilon$}^\dag_3)Z(\Lambda_2,m_2,r)\nonumber\\&&+\frac{1}{3}S(\hat{\mbox{\boldmath$r$}},\mbox{\boldmath$\epsilon$}_2,
\mbox{\boldmath$\epsilon$}^\dag_3)T(\Lambda_2,m_2,r)\bigg],
\\
V^{\mathrm{Cross}}_ \eta&=& -\frac{g^2}{f_\pi^2}
\bigg[\frac{1}{3}(\mbox{\boldmath$\epsilon$}_2\cdot
\mbox{\boldmath$\epsilon$}^\dag_3)Z(\Lambda_3,m_3,r)\nonumber\\&&
+\frac{1}{3}S(\hat{\mbox{\boldmath$r$}},\mbox{\boldmath$\epsilon$}_2,
\mbox{\boldmath$\epsilon$}^\dag_3)T(\Lambda_3,m_3,r)],\\
V^{\mathrm{Direct}}_\sigma&=& -g_s^2(\mbox{\boldmath$\epsilon$}_2\cdot \mbox{\boldmath$\epsilon$}^\dag_4 ) Y(\Lambda,m_\sigma,r),\\
V^{\mathrm{Direct}}_{\rho}&=& -\frac{1}{2}\beta^2g_V^2 (\mbox{\boldmath$\epsilon$}_2\cdot \mbox{\boldmath$\epsilon$}^\dag_4)Y(\Lambda,m_\rho,r),\\
V^{\mathrm{Cross}}_{\rho}&=&2\lambda^2g_V^2\bigg[\frac{2}{3}(\mbox{\boldmath$\epsilon$}_2\cdot
\mbox{\boldmath$\epsilon$}^\dag_3)Z(\Lambda_0,m_0,r)\nonumber\\&&-\frac{1}{3}S(\hat{\mbox{\boldmath$r$}},\mbox{\boldmath$\epsilon$}_2,
\mbox{\boldmath$\epsilon$}^\dag_3)T(\Lambda_0,m_0,r)\bigg],\\
V^{\mathrm{Direct}}_{\omega} &=& -\frac{1}{2}\beta^2g_V^2 (\mbox{\boldmath$\epsilon$}_2\cdot \mbox{\boldmath$\epsilon$}^\dag_4)Y(\Lambda,m_\omega,r),\\
V^{\mathrm{Cross}}_{\omega}&=& 2 \lambda^2g_V^2\bigg[\frac{2}{3}
(\mbox{\boldmath$\epsilon$}_2\cdot
\mbox{\boldmath$\epsilon$}^\dag_3)Z(\Lambda_1,m_1,r)\nonumber\\&&-\frac{1}{3}
S(\hat{\mbox{\boldmath$r$}},\mbox{\boldmath$\epsilon$}_2,\mbox{\boldmath$\epsilon$}^\dag_3)
T(\Lambda_1,m_1,r)\bigg].
\end{eqnarray}
In the above expressions, we define
\begin{eqnarray*}
\Lambda_2^2&=&\Lambda^2-(m_{B^*}-m_B)^2,\quad
m_2^2=m_\pi^2-(m_{B^*}-m_B)^2,\\
\Lambda_3^2&=&\Lambda^2-(m_{B^*}-m_B)^2,\quad
m_3^2=m_\eta^2-(m_{B^*}-m_B)^2,\\
\Lambda_0^2&=&\Lambda^2-(m_{B^*}-m_B)^2,\quad
m_0^2=m_\rho^2-(m_{B^*}-m_B)^2,\\
\Lambda_1^2&=&\Lambda^2-(m_{B^*}-m_B)^2,\quad
m_1^2=m_\omega^2-(m_{B^*}-m_B)^2,
\end{eqnarray*}
and
$S(\hat{\mbox{\boldmath$r$}},\mathbf{a},\mathbf{b})=3(\hat{\mbox{\boldmath$r$}}\cdot\mathbf{a})
(\hat{\mbox{\boldmath$r$}}\cdot\mathbf{b})-\mathbf{a}\cdot\mathbf{b}$.
Additionally, functions $Y(\Lambda,m,r)$, $Z(\Lambda,m,r)$ and
$T(\Lambda,m,r)$ are defined as
\begin{eqnarray}
Y(\Lambda,m_E,r) &=& \frac{1}{4\pi r}(e^{-m_E\,r}-e^{-\Lambda r})-\frac{\Lambda^2-m_E^2}{8\pi \Lambda }e^{-\Lambda r},\label{m1}\\
Z(\Lambda,m_E,r) &=& \bigtriangledown^2Y(\Lambda,m_E,r) = \frac{1}{r^2} \frac{\partial}{\partial r}r^2 \frac{\partial}{\partial r}Y(\Lambda,m_E,r),\\
T(\Lambda,m_E,r) &=&  r\frac{\partial}{\partial
r}\frac{1}{r}\frac{\partial}{\partial
r}Y(\Lambda,m_E,r).\label{m3}
\end{eqnarray}
In Eqs. (\ref{h1})-(\ref{h2}), $c=+1$ corresponds to the
$Z_{B\bar{B}^*}^{(T)}$ and $Z_{B\bar{B}^*}^{(S)}$ states including
these two charged $Z_b$ states observed by Belle collaboration
while taking $c=-1$ corresponds to the
${Z_{B\bar{B}^*}^{(T)}}^\prime$ and
${Z_{B\bar{B}^*}^{(S)}}^\prime$ states which are partner states of
X(3872).

As indicated in Eq. (\ref{11}), we consider both S-wave and D-wave
interactions between the $B$ and $\bar{B}^*$ mesons. Finally the
total effective potential can be obtained by making the
replacement in the subpotentials
\begin{eqnarray}
\left . \begin{array}{c}
(\mbox{\boldmath$\epsilon$}_2\cdot \mbox{\boldmath$\epsilon$}^\dag_3)\\
(\mbox{\boldmath$\epsilon$}_2\cdot
\mbox{\boldmath$\epsilon$}^\dag_4 )
\end{array}\right\} \rightarrowtail \left(\begin{array}{cc}
1&0\\
0&1\\
\end{array}\right),\quad
S(\hat{\mbox{\boldmath$r$}},\mbox{\boldmath$\epsilon$}_2,
\mbox{\boldmath$\epsilon$}^\dag_3) \rightarrowtail
\left(\begin{array}{cc}
0&-\sqrt{2}\\
-\sqrt{2}&1\\
\end{array}\right),\nonumber
\end{eqnarray}
which results in the total effective potential of the $B\bar{B}^*$
system, i.e, a two by two matrix.

The effective potential of the $D\bar{D}^*$ system is similar to
that of $B\bar{B}^*$ system. The $\eta$, $\sigma$, $\rho$ and
$\omega$ meson exchange potentials of $D\bar{D}^*$ system can be
easily obtained by replacing the parameters for the $B\bar{B}^*$
system with the ones for $D\bar{D}^*$ system. Since the mass gap
of $m_D^*$ and $m_D$ is larger than the mass of $\pi$, which is
different from the case of the $B\bar{B}^*$ system, the $\pi$
exchange potential of the $D\bar{D}^*$ system is
\cite{Liu:2008fh,Liu:2008tn}
\begin{eqnarray}
V^{\mathrm{Cross}}_\pi&=&
-\frac{g^2}{f_\pi^2}\bigg[\frac{1}{3}(\mbox{\boldmath$\epsilon$}_2\cdot
\mbox{\boldmath$\epsilon$}^\dag_3)Z^{DD^*}_\pi(\Lambda_4,m_4,r)\nonumber\\&&+\frac{1}{3}S(\hat{\mbox{\boldmath$r$}},\mbox{\boldmath$\epsilon$}_2,
\mbox{\boldmath$\epsilon$}^\dag_3)T^{DD^*}_\pi(\Lambda_4,m_4,r)\bigg],
\end{eqnarray}
where
\begin{eqnarray}
Y^{DD^*}_\pi(\Lambda_4,m_4,r) &=& \frac{1}{4\pi r}\bigg(-e^{-\Lambda_4 r}-\frac{r(\Lambda_4^2+m_4^2)}{2\Lambda_4}e^{-\Lambda_4 r}\nonumber\\
&&+\cos({m_4r})\bigg),\\
Z^{DD^*}_\pi(\Lambda_4,m_4,r) &=& \bigtriangledown^2Y^{DD^*}_\pi(\Lambda_4,m_4,r) = \frac{1}{r^2} \frac{\partial}{\partial r}r^2 \frac{\partial}{\partial r}\nonumber\\
&&\times Y^{DD^*}_\pi(\Lambda_4,m_4,r),\\
T^{DD^*}_\pi(\Lambda_4,m_4,r) &=&  r\frac{\partial}{\partial
r}\frac{1}{r}\frac{\partial}{\partial
r}Y^{DD^*}_\pi(\Lambda_4,m_4,r).
\end{eqnarray}
In the present case, the parameters $\Lambda_4$ and $m_4$ are
defined as
\begin{eqnarray}
\Lambda_4 &=& \sqrt{\Lambda^2-(m_{D^*}-m_D)^2},\\
m_4 &=& \sqrt{(m_{D^*}-m_D)^2-m_\pi^2}.
\end{eqnarray}

\subsubsection{The $B^*\bar{B}^*$ system}

For the isoscalar and isovector $B^*\bar{B}^*$ systems, the
general expressions of the total effective potentials are
\begin{eqnarray}
\mathcal{V}^{{Z_{B^*\bar{B}^*}^{(T)}}^{(\prime)}}&=&
W_\sigma-\frac{1}{2}W_\rho
+\frac{1}{2}W_\omega-\frac{1}{2}W_\pi+\frac{1}{6}W_\eta,\label{h11}\\
\mathcal{V}^{{Z_{B^*\bar{B}^*}^{(S)}}^{(\prime)}}&=&
W_\sigma+\frac{3}{2}W_\rho
+\frac{1}{2}V_\omega+\frac{3}{2}W_\pi+\frac{1}{6}W_\eta,\label{h21}
\end{eqnarray}
respectively, where the $\pi$, $\eta$, $\sigma$, $\rho$ and
$\omega$ meson exchanges can contribute to the effective
potentials. The corresponding subpotentials are expressed as
\begin{eqnarray}
W_\pi&=&
-\frac{g^2}{f_\pi^2}\bigg[\frac{1}{3}(\mbox{\boldmath$\epsilon$}_1\times
\mbox{\boldmath$\epsilon$}^\dag_3)\cdot(\mbox{\boldmath$\epsilon$}_2\times
\mbox{\boldmath$\epsilon$}^\dag_4)Z(\Lambda,m_\pi,r)\nonumber\\&&+\frac{1}{3}S(\hat{\mbox{\boldmath$r$}},\mbox{\boldmath$\epsilon$}_1\times
\mbox{\boldmath$\epsilon$}^\dag_3,
\mbox{\boldmath$\epsilon$}_2\times
\mbox{\boldmath$\epsilon$}^\dag_4)T(\Lambda,m_\pi,r)\bigg],
\\
W_\eta&=& -\frac{g^2}{f_\pi^2}
\bigg[\frac{1}{3}(\mbox{\boldmath$\epsilon$}_1\times
\mbox{\boldmath$\epsilon$}^\dag_3)\cdot(\mbox{\boldmath$\epsilon$}_2\times
\mbox{\boldmath$\epsilon$}^\dag_4)Z(\Lambda,m_\eta,r)\nonumber\\&&
+\frac{1}{3}S(\hat{\mbox{\boldmath$r$}},\mbox{\boldmath$\epsilon$}_1\times
\mbox{\boldmath$\epsilon$}^\dag_3,
\mbox{\boldmath$\epsilon$}_2\times \mbox{\boldmath$\epsilon$}^\dag_4)T(\Lambda,m_\eta,r)\bigg],\\
W_\sigma&=& -g_s^2(\mbox{\boldmath$\epsilon$}_1\cdot \mbox{\boldmath$\epsilon$}^\dag_3 ) (\mbox{\boldmath$\epsilon$}_2\cdot \mbox{\boldmath$\epsilon$}^\dag_4 ) Y(\Lambda,m_\sigma,r),\\
W_{\rho}&=& -\frac{1}{4} \bigg\{2\beta^2g_V^2(\mbox{\boldmath$\epsilon$}_1\cdot \mbox{\boldmath$\epsilon$}^\dag_3)(\mbox{\boldmath$\epsilon$}_2\cdot \mbox{\boldmath$\epsilon$}^\dag_4)Y(\Lambda,m_\rho,r)\nonumber\\
&&-8\lambda^2g_V^2\bigg[\frac{2}{3}(\mbox{\boldmath$\epsilon$}_1\times \mbox{\boldmath$\epsilon$}^\dag_3)\cdot(\mbox{\boldmath$\epsilon$}_2\times \mbox{\boldmath$\epsilon$}^\dag_4)Z(\Lambda,m_\rho,r)\nonumber\\
&&-\frac{1}{3}S(\hat{\mbox{\boldmath$r$}},\mbox{\boldmath$\epsilon$}_1\times \mbox{\boldmath$\epsilon$}^\dag_3,\mbox{\boldmath$\epsilon$}_2\times \mbox{\boldmath$\epsilon$}^\dag_4)T(\Lambda,m_\rho,r)\bigg]\bigg\},\\
W_{\omega} &=& -\frac{1}{4} \bigg\{2\beta^2g_V^2(\mbox{\boldmath$\epsilon$}_1\cdot \mbox{\boldmath$\epsilon$}^\dag_3)(\mbox{\boldmath$\epsilon$}_2\cdot \mbox{\boldmath$\epsilon$}^\dag_4)Y(\Lambda,m_\omega,r)\nonumber\\
&&-8\lambda^2g_V^2\bigg[\frac{2}{3}(\mbox{\boldmath$\epsilon$}_1\times \mbox{\boldmath$\epsilon$}^\dag_3)\cdot(\mbox{\boldmath$\epsilon$}_2\times \mbox{\boldmath$\epsilon$}^\dag_4)Z(\Lambda,m_\omega,r)\nonumber\\
&&-\frac{1}{3}S(\hat{\mbox{\boldmath$r$}},\mbox{\boldmath$\epsilon$}_1\times
\mbox{\boldmath$\epsilon$}^\dag_3,\mbox{\boldmath$\epsilon$}_2\times
\mbox{\boldmath$\epsilon$}^\dag_4)T(\Lambda,m_\omega,r)\bigg]\bigg\}.
\end{eqnarray}
Here, the definitions of $Y(\Lambda,m,r)$,
$Z(\Lambda,m,r),T(\Lambda,m,r)$ and $S(\hat{\mbox{\boldmath
$r$}},\mathbf{a},\mathbf{b})$ are given in Sec. \ref{bbs}.

In this work, we consider both S-wave and D-wave interactions
between the $B^*$ and $\bar{B}^*$ mesons, which are illustrated in
Eq. (\ref{11}). Thus, the total effective potential of the
$B^*\bar{B}^*$ with $J=0,1,2$ is $2\times 2$, $3\times 3$,
$4\times 4$ matrices, which can be obtained by replacing the
corresponding terms in the subpotentials, i.e.,
\begin{eqnarray}
&&(\mbox{\boldmath$\epsilon$}_1\cdot
\mbox{\boldmath$\epsilon$}^\dag_3)(\mbox{\boldmath$\epsilon$}_2\cdot
\mbox{\boldmath$\epsilon$}^\dag_4)
 \rightarrowtail \left(\begin{array}{ccc}
1&0\\
0&1\\
\end{array}\right),\\
&&(\mbox{\boldmath$\epsilon$}_1\times
\mbox{\boldmath$\epsilon$}^\dag_3)\cdot(\mbox{\boldmath$\epsilon$}_2\times
\mbox{\boldmath$\epsilon$}^\dag_4) \rightarrowtail
\left(\begin{array}{ccc}
2&0\\
0&-1\\
\end{array}\right),
\\
&&S(\hat{\mbox{\boldmath$r$}},\mbox{\boldmath$\epsilon$}_1\times
\mbox{\boldmath$\epsilon$}^\dag_3,
\mbox{\boldmath$\epsilon$}_2\times
\mbox{\boldmath$\epsilon$}^\dag_4) \rightarrowtail
\left(\begin{array}{ccc}
0&\sqrt{2}\\
\sqrt{2}&2\\
\end{array}\right)
\end{eqnarray}
for the $B^*\bar{B}^*$ states with $J=0$,
\begin{eqnarray}
&&(\mbox{\boldmath$\epsilon$}_1\cdot
\mbox{\boldmath$\epsilon$}^\dag_3)(\mbox{\boldmath$\epsilon$}_2\cdot
\mbox{\boldmath$\epsilon$}^\dag_4)
 \rightarrowtail \left(\begin{array}{ccc}
1&0&0\\
0&1&0\\
0&0&1\\
\end{array}\right),\\
&&(\mbox{\boldmath$\epsilon$}_1\times
\mbox{\boldmath$\epsilon$}^\dag_3)\cdot(\mbox{\boldmath$\epsilon$}_2\times
\mbox{\boldmath$\epsilon$}^\dag_4) \rightarrowtail
\left(\begin{array}{ccc}
1&0&0\\
0&1&0\\
0&0&-1\\
\end{array}\right),
\\
&&S(\hat{\mbox{\boldmath$r$}},\mbox{\boldmath$\epsilon$}_1\times
\mbox{\boldmath$\epsilon$}^\dag_3,
\mbox{\boldmath$\epsilon$}_2\times
\mbox{\boldmath$\epsilon$}^\dag_4) \rightarrowtail
\left(\begin{array}{ccc}
0&-\sqrt{2}&0\\
-\sqrt{2}&1&0\\
0&0&1\\
\end{array}\right)
\end{eqnarray}
for the $B^*\bar{B}^*$ states with $J=1$, and
\begin{eqnarray}
&&(\mbox{\boldmath$\epsilon$}_1\cdot
\mbox{\boldmath$\epsilon$}^\dag_3)(\mbox{\boldmath$\epsilon$}_2\cdot
\mbox{\boldmath$\epsilon$}^\dag_4)
 \rightarrowtail \left(\begin{array}{cccc}
1&0&0&0\\
0&1&0&0\\
0&0&1&0\\
0&0&0&1
\end{array}\right),\\
&&(\mbox{\boldmath$\epsilon$}_1\times
\mbox{\boldmath$\epsilon$}^\dag_3)\cdot(\mbox{\boldmath$\epsilon$}_2\times
\mbox{\boldmath$\epsilon$}^\dag_4) \rightarrowtail
\left(\begin{array}{cccc}
-1&0&0&0\\
0&2&0&0\\
0&0&1&0\\
0&0&0&-1
\end{array}\right),
\\
&&S(\hat{\mbox{\boldmath$r$}},\mbox{\boldmath$\epsilon$}_1\times
\mbox{\boldmath$\epsilon$}^\dag_3,
\mbox{\boldmath$\epsilon$}_2\times
\mbox{\boldmath$\epsilon$}^\dag_4) \rightarrowtail
\left(\begin{array}{cccc}
0&\sqrt{\frac{2}{5}}&0&-\sqrt{\frac{14}{5}}\\
\sqrt{\frac{2}{5}}&0&0&-\frac{2}{\sqrt{7}}\\
0&0&-1&0\\
-\sqrt{\frac{14}{5}}&{-\frac{2}{\sqrt{7}}}&0&-\frac{3}{7}
\end{array}\right)\nonumber\\
\end{eqnarray}
for the $B^*\bar{B}^*$ states with $J=2$.

The potentials of the $D^*\bar{D}^*$ system and $B^*\bar{B}^*$
system have the same form. We only need to replace the parameters
for the $B^*\bar{B}^*$ system with the ones for the $D^*\bar{D}^*$
system.

\section{Numerical results}\label{sec3}

With the obtained effective potentials, we can find the bound
state solution by solving the coupled-channel Schr\"odinger
equation. Corresponding to the systems in Eqs.
(\ref{h1})-(\ref{h2}), the kinetic terms for the
${Z_{B\bar{B}^*}^{(\alpha)}}^\prime$ and
$Z_{B^*\bar{B}^*}^{(\alpha)}[\mathrm{J}]$ ($\mathrm{J}=0,1,2$)
systems are
\begin{eqnarray}
    K_{{Z_{B\bar{B}^*}^{(\alpha)}}^\prime}&=&\mathrm{diag}\Bigg(-\frac{\triangle}{2\tilde{m}_1},~
    -\frac{\triangle_2}{2\tilde{m}_1}\Bigg),\\
    K_{Z_{B^*\bar{B}^*}^{(\alpha)}[0]}&=&\mathrm{diag}\Bigg(-\frac{\triangle}{2\tilde{m}_2},~
    -\frac{\triangle_2}{2\tilde{m}_2}\Bigg),\\
    K_{Z_{B^*\bar{B}^*}^{(\alpha)}[1]}&=&\mathrm{diag}\Bigg(-\frac{\triangle}{2\tilde{m}_2},~
    -\frac{\triangle_2}{2\tilde{m}_2},~
    -\frac{\triangle_2}{2\tilde{m}_2}\Bigg),\\
K_{Z_{B^*\bar{B}^*}^{(\alpha)}[2]}&=&\mathrm{diag}\Bigg(-\frac{\triangle}{2\tilde{m}_2},~
    -\frac{\triangle_2}{2\tilde{m}_2},~
    -\frac{\triangle_2}{2\tilde{m}_2},~
    -\frac{\triangle_2}{2\tilde{m}_2}\Bigg),
\end{eqnarray}
respectively. Here,
$\triangle=\frac{1}{r^2}\frac{\partial}{\partial
r}r^2~\frac{\partial}{\partial r}$, $\triangle_2=\triangle
-{6\over{r^2}}$. $\tilde{m}_1=m_Bm_{B^{*}}/(m_B+m_{B^{*}})$ and
$\tilde{m}_2=m_{B^*}/2$ are the reduced masses of the
$Z_{b1}^{(i)}$ and $Z_{b2}^{(i)}$ systems, where $m_B$ and
$m_{B^{*}}$ denote the masses of the pseudoscalar and vector
bottom mesons \cite{Nakamura:2010zzi}, respectively. Of course,
the kinematic terms for the $D\bar{D}^*$ and $D^*\bar{D}^*$
systems are of the same forms as those for the $B\bar{B}^*$ and
$B^*\bar{B}^*$ systems, where we replace the mass of $D^{(*)}$
with that of $B^{(*)}$.

In this work, the FESSDE
program~\cite{Abrashkevich1995,Abrashkevich1998} is adopted to
produce the numerical values for the binding energy and the
relevant root-mean-square $r$ with the variation of the cutoff in
the region of $0.8\leq\Lambda\leq5$~GeV. Moreover, we also use
MATSCE \cite{matscs}, a MATLAB package for solving coupled-channel
Schr\"{o}dinger equation, to perform an independent cross-check.

Throughout this work, we will first present the numerical results
of the obtained bound state solutions when all types of the
one-meson-exchange (OME) potentials are included. The
one-pion-exchange force contributes to the long range interaction
between the heavy meson pair, which is clear and well-known. In
contrast, the scalar and vector meson exchanges are used to mimic
the intermediate and short range interaction between the heavy
mesons, which are not determined very precisely. In order to find
out whether the existence of the possible bound molecular states
is sensitive to the details of the short-range interaction, we
will also study the case when only the one-pion-exchange (OPE)
contribution is considered. In the following illustration, we use
OME and OPE to distinguish such two cases. If the OPE force alone
is strong enough to form a loosely bound state, such a case is
particularly interesting phenomenologically.

\subsection{The $B\bar{B}^*$ and $D\bar{D}^*$ systems}

In the following, we first present the numerical results for the
$Z_{B\bar{B}^*}^{(T)}$ and $Z_{B\bar{B}^*}^{(S)}$ states where
$Z_{B\bar{B}^*}^{(T)}$ corresponds to $Z_b(10610)$ observed by
Belle \cite{Collaboration:2011gj}. As shown in Table \ref{wave},
there exist two systems with $c=-1$ and $C=+1$ in the flavor wave
functions, where are marked as ${Z_{B\bar{B}^*}^{(T)}}^\prime$ and
${Z_{B\bar{B}^*}^{(S)}}^\prime$.

\begin{enumerate}
\item{In Table \ref{BBS-1}, we present the numerical results of
the obtained bound state solutions in both OME and OPE cases. We
find the bound state solutions for the two isoscalar
$Z_{B\bar{B}^*}^{(S)}$ and ${Z_{B\bar{B}^*}^{(S)}}^\prime$ with
reasonable $\Lambda$ values ($\Lambda\sim 1$ GeV), which indicates
the existence of the $Z_{B\bar{B}^*}^{(S)}$ and
${Z_{B\bar{B}^*}^{(S)}}^\prime$ molecular states.}

\item{For the $Z_{B\bar{B}^*}^{(T)}$ state, we also find the bound
state solution with $\Lambda$ around 2.2 GeV. Our result shows
that $Z_{B\bar{B}^*}^{(T)}$ could be as a molecular state with a
very shallow binding energy. In addition, its binding energy is
not strongly dependent on $\Lambda$. Thus, it is quite natural to
interpret $Z_b(10610)$ as a $B\bar{B}^*$ molecular state with
isospin $I=1$.}

\item{For the ${Z_{B\bar{B}^*}^{(T)}}^\prime$ system, the bound
state solution can be found in the region $\Lambda>4.7$ GeV. To
some extent, the value of $\Lambda$ for the
${Z_{B\bar{B}^*}^{(T)}}^\prime$ seems a little large compared to 1
GeV. }

\item{We also discuss the case when we only consider the OPE
potential. For the \{$Z_{B\bar{B}^*}^{(T)}$,
${Z_{B\bar{B}^*}^{(T)}}^\prime$, ${Z_{B\bar{B}^*}^{(S)}}^\prime$\}
or $Z_{B\bar{B}^*}^{(S)}$, we need to decrease or increase the
$\Lambda$ value to obtain the same binding energy as that from
OME. The one pion meson exchange potential indeed plays the
crucial role in the formation of the $BB^*$ bound states. }
\end{enumerate}

\renewcommand{\arraystretch}{1.6}
\begin{table}[hbt]
\caption{The obtained bound state solutions (binding energy $E$
and root-mean-square radius $r_{\mathrm{RMS}}$) for the
$B\bar{B}^*$ systems. Here, we discuss two situations, i.e.,
including all one meson exchange (OME) contribution and only
considering one pion exchange (OPE) potential. \label{BBS-1}}
\begin{tabular}{c|cccc|cccc}\toprule[1pt]
&&\multicolumn{3}{c|}{OME}&\multicolumn{3}{c}{OPE}\\\midrule[1pt]
$I^G(J^{PC})$ & State& $\Lambda$ & $E$ (MeV)& $r_{\mathrm{RMS}}$
(fm) & $\Lambda$ & $E$ (MeV)& $r_{\mathrm{RMS}}$ (fm)
\\\midrule[1pt]
\multirow{3}*{$1^+(1^{+})$}&\multirow{3}*{$Z_{B\bar{B}^*}^{(T)}$}&2.1&-0.22&3.05&2.2&-8.69&0.62\\
                          &&2.3&-1.64&1.31&2.4&-20.29&0.47\\
                          &&2.5&-4.74&0.84&2.6&-38.54&0.36\\\midrule[1pt]
\multirow{3}*{$1^-(1^{+})$}&\multirow{3}*{${Z_{B\bar{B}^*}^{(T)}}^\prime$}&4.9&-0.14&3.64&4.5&-17.79&0.56\\
            &             &5.0&-0.41&2.45&4.6&-22.65&0.52\\
            &             &5.1&-0.85&1.80&4.7&-28.29&0.48\\\midrule[1pt]
\multirow{3}*{$0^-(1^{+-})$}&\multirow{3}*{${Z_{B\bar{B}^*}^{(S)}}$}&1.0&-0.28&3.35&1.8&-10.09&0.96\\
                &           &1.05&-1.81&1.71&1.9&-15.11&0.84\\
                &           &1.1&-5.36&1.18&2.0&-21.53&0.76\\
\midrule[1pt]
\multirow{3}*{$0^+(1^{++})$}&\multirow{3}*{${Z_{B\bar{B}^*}^{(S)}}^\prime$}&0.8&-0.95&1.84&1.0&-7.68&0.82\\
                                   &&0.9&-6.81&0.91&1.1&-15.30&0.65\\
                                   &&1.0&-19.92&0.65&1.2&-26.53&0.53\\
\bottomrule[1pt]
\end{tabular}
\end{table}

We extend the formalism in Sec. \ref{sec2} to study the
$D\bar{D}^*$ systems. As shown in Table \ref{BBS-2}, we can
exclude the existence of the $Z_{D\bar{D}^*}^{(T)}$ and
${Z_{D\bar{D}^*}^{(T)}}^\prime$ since we do not find any bound
state solution for the $Z_{D\bar{D}^*}^{(T)}$ and
${Z_{D\bar{D}^*}^{(T)}}^\prime$ states. For the two isoscalar
$Z_{D\bar{D}^*}^{(S)}$ and ${Z_{D\bar{D}^*}^{(S)}}^\prime$, there
exist loosely bound states with reasonable $\Lambda$ values. If
only considering the OPE exchange potential, we notice: (1) the
bound state solution of the $Z_{D\bar{D}^*}^{(T)}$ appears when
$\Lambda\sim 4.6$ GeV, which largely deviates from 1 GeV; (2)
there still does not exist any bound state solution for
${Z_{D\bar{D}^*}^{(T)}}^\prime$; (3) for $Z_{D\bar{D}^*}^{(S)}$
and ${Z_{D\bar{D}^*}^{(S)}}^\prime$, $\Lambda$ becomes larger in
order to find the bound state solution. The comparison between the
OME and OPE results also reflects the importance of one pion
exchange in the $D\bar{D}^*$ systems. We need to specify that
${Z_{D\bar{D}^*}^{(S)}}^\prime$ with $0^+(1^{++})$ directly
corresponds to the observed $X(3872)$ \cite{Choi:2003ue}.

The BaBar Collaboration measured the radiative decay of $X(3872)$ and
found a ratio of $B(X(3872)\to \psi(2S)\gamma)/B(X(3872)\to
J/\psi\gamma)=3.4\pm1.4$ \cite{:2008rn}, which contradicts the
prediction with a purely $D\bar{D}^*$ molecular assignment to
$X(3872)$ \cite{Swanson:2004pp}. However, very recently Belle
reported a new measurement of the radiative decay of $X(3872)$,
where only the decay mode $X(3872)\to J/\psi\gamma$ was observed
and the upper limit $B(X(3872)\to \psi(2S)\gamma)/B(X(3872)\to
J/\psi\gamma)<2.1$ was given \cite{Bhardwaj:2011dj}. The
inconsistence between the Belle and BaBar results indicate that the
study of $X(3872)$ is still an important research topic. Our
numerical results suggest that the mass of the loosely bound
molecular state ${Z_{D\bar{D}^*}^{(S)}}^\prime$ is consistent with
that of $X(3872)$. The assignment of $X(3872)$ as a molecular
candidate is still very attractive.

\renewcommand{\arraystretch}{1.6}
\begin{table}[hbt]
\caption{The obtained bound state solutions (binding energy $E$ and root-mean-square radius $r_{\mathrm{RMS}}$) for $D\bar{D}^*$ systems.
Here, we discuss two situations, i.e., including all one meson exchange (OME) contribution and only considering one pion exchange (OPE) potential to $B\bar{B}^*$ systems. \label{BBS-2}}
\begin{tabular}{c|cccc|cccc}\toprule[1pt]
&&\multicolumn{3}{c|}{OME}&\multicolumn{3}{c}{OPE}\\\midrule[1pt]
$I^G(J^{PC})$ & State& $\Lambda$ & $E$ (MeV)& $r_{\mathrm{RMS}}$ (fm) & $\Lambda$ & $E$ (MeV)& $r_{\mathrm{RMS}}$ (fm) \\\midrule[1pt]
\multirow{4}*{$1^+(1^{+-})$}&\multirow{4}*{$Z_{D\bar{D}^*}^{(T)}$}&\multirow{4}*{-}&\multirow{4}*{-}&\multirow{4}*{-}
                          &4.6&-0.85&1.46\\
                          &&&&&4.7&-3.42&1.17\\
                          &&&&&4.8&-7.18&0.93\\
                          &&&&&4.9&-12.40&0.75\\
                          \midrule[1pt]
\multirow{1}*{$1^-(1^{++})$}&\multirow{1}*{${Z_{D\bar{D}^*}^{(T)}}^\prime$}&-&-&-&-&-&-\\
                  \midrule[1pt]
\multirow{4}*{$0^-(1^{+-})$}&\multirow{4}*{${Z_{D\bar{D}^*}^{(S)}}$}&1.3&-&-&3.4&-0.11&1.74\\
                &
		&1.4&-1.56&1.61&3.5&-2.03&1.50\\
                &
		&1.5&-12.95&0.98&3.6&-4.79&1.26\\
                &
		&1.6&-35.73&0.69&3.7&-9.62&1.06\\
\midrule[1pt]
\multirow{4}*{$0^+(1^{++})$}&\multirow{4}*{${Z_{D\bar{D}^*}^{(S)}}^\prime$}&1.1&-0.61&&1.7&-3.01&1.37\\
                                   &&1.2&-4.42&1.38&1.8&-7.41&1.06\\
                                   &&1.3&-11.78&1.05&1.9&-14.15&0.84\\
                                   &&1.4&-21.88&0.86&2&-23.82&0.68\\
\bottomrule[1pt]
\end{tabular}
\end{table}

\subsection{The $B^*\bar{B}^*$ and $D^*\bar{D}^*$ systems}

The numerical results of the $B^*\bar{B}^*$ systems are presented
in Table \ref{BSBS-1}, which include the obtained binding energy
and the corresponding root-mean-square radius. We find the bound
state solution for all the $B^*\bar{B}^*$ states with reasonable
$\Lambda$ values:

\begin{enumerate}
\item{A loosely bound state exists for $Z_{B^*\bar{B}*}^{(T)}[1]$
corresponding to the observed $Z_b(10650)$ with $\Lambda$ slightly
above 2 GeV. Only considering the OPE potential, the obtained
binding energy becomes deeper with the same $\Lambda$ value.  }

\item{In addition, the $B^*\bar{B}^*$ can form loosely bound
molecular states $Z_{B^*\bar{B}*}^{(S)}[0]$,
$Z_{B^*\bar{B}*}^{(T)}[0]$, $Z_{B^*\bar{B}*}^{(S)}[1]$ and
$Z_{B^*\bar{B}*}^{(S)}[2]$ with very reasonable $\Lambda$ values.
Comparing the results between OME and OPE cases, one notices again
that the one pion exchange indeed is very important to form the
$B^*\bar{B}^*$ bound state.}

\item{For the $Z_{B^*\bar{B}*}^{(T)}[2]$ state, the existence of
the loosely bound state requires the value of $\Lambda$ around 4.4
GeV. }

\end{enumerate}

\renewcommand{\arraystretch}{1.6}
\begin{table}[hbtp]
\caption{The obtained bound state solutions (binding energy $E$ and root-mean-square radius $r_{\mathrm{RMS}}$) for $B^*\bar{B}^*$ systems.
Here, we discuss two situations, i.e., including all one meson exchange (OME) contribution and only considering one pion exchange (OPE) potential to $B\bar{B}^*$systems. \label{BSBS-1}}
\begin{tabular}{c|cccc|cccc}\toprule[1pt]
&&\multicolumn{3}{c|}{OME}&\multicolumn{3}{c}{OPE}\\\midrule[1pt]
$I^G(J^{PC})$ & State& $\Lambda$ & $E$ (MeV)& $r_{\mathrm{RMS}}$ (fm)& $\Lambda$ & $E$ (MeV)& $r_{\mathrm{RMS}}$ (fm)\\\midrule[1pt]
\multirow{4}*{$1^{+}(0^+)$}&\multirow{4}*{$Z_{B^*B^*}^{(T)}[0]$}&1.2&-&-&1&-&-\\
                        &&1.4&-1.44&1.24&1.2&-0.32&1.53\\
                        &&1.6&-6.16&0.77&1.4&-5.69&0.78\\
                        &&1.8&-15.15&0.54&1.6&-18.82&0.50\\\midrule[1pt]
\multirow{4}*{$0^{-}(0^{+-})$}&\multirow{4}*{$Z_{B^*B^*}^{(S)}[0]$}&0.9&-&-&1&-&-\\
            &&             1&-0.81&2.11&1.2&-0.52&2.76\\
            &&
	    1.1&-9.98&1.02&1.4&-5.74&1.12\\
            &&
	    1.2&-35.16&0.70&1.6&-20.92&0.77\\
\midrule[1pt]
\multirow{4}*{$1^{+}(1^+)$}&\multirow{4}*{$Z_{B^*B^*}^{(T)}[1]$}&
		2.2&-0.81&1.38&2&-2.17&1.15\\
                &&2.4&-3.31&0.95&2.2&-8.01&0.68\\
                &&
		2.6&-7.80&0.68&2.4&-19.00&0.48\\
                &&
		2.8&-14.94&0.52&2.6&-36.36&0.38\\
                \midrule[1pt]
\multirow{4}*{$0^{-}(1^{+-})$}&\multirow{4}*{$Z_{B^*B^*}^{(S)}[1]$}&1.&-0.01&2.07&1.4&-0.51&1.90\\
                                   &&1.1&-5.50&1.17&1.6&3.65&-1.32\\
                                   &&1.2&-21.76&-0.75&1.8&-10.26&0.96\\
                                   &&1.3&-53.68&0.55&2.0&-21.81&0.75\\\midrule[1pt]
\multirow{4}*{$1^{+}(2^+)$}&\multirow{4}*{$Z_{B^*B^*}^{(T)}[2]$}&4.4&-0.44&1.59&3.6&-2.82&1.12\\
                                   &&4.6&-1.59&1.28&3.8&-6.21&0.85\\
                                   &&4.8&-3.42&1.01&4.0&-11.41&0.68\\
                                   &&5.&-6.16&0.81&4.2&-18.77&0.57\\\midrule[1pt]
\multirow{3}*{$0^{-}(2^{+-})$}&\multirow{3}*{$Z_{B^*B^*}^{(S)}[2]$}
                                   &0.8&-2.33&1.32&0.8&-1.81&1.48\\
                                   &&0.9&-10.45&0.84&0.9&-5.64&1.01\\
                                   &&1.0&-27.14&0.63&1.0&-12.28&0.76\\
\bottomrule[1pt]

\end{tabular}
\end{table}

In the following, we also present the numerical results for the
$D^*\bar{D}^*$ systems in Table \ref{DSDS-1}. Our calculation
indicates:
\begin{enumerate}

\item{We find the bound state solutions for the three isoscalar
states $Z_{D^*\bar{D}^*}^{(S)}[0]$, $Z_{D^*\bar{D}^*}^{(S)}[1]$
and $Z_{D^*\bar{D}^*}^{(S)}[2]$, where the corresponding $\Lambda$
is around 1 GeV. If only considering the OPE contribution for the
$Z_{D^*\bar{D}^*}^{(S)}[0]$, $Z_{D^*\bar{D}^*}^{(S)}[1]$ states,
we need to largely increase $\Lambda$ value in order to obtain a
loosely bound state. Here, either $Z_{D^*\bar{D}^*}^{(S)}[0]$ or
$Z_{D^*\bar{D}^*}^{(S)}[2]$ could correspond to the observed
$Y(3930)$ by Belle \cite{Abe:2004zs} and BaBar
\cite{Aubert:2007vj}, which is consistent with the conclusion in
Ref. \cite{Liu:2009ei}.}

\item{There does not exist the bound state
$Z_{D^*\bar{D}^*}^{(T)}[2]$. The value of $\Lambda$ is about 3.6
GeV in order to form a bound state $Z_{D^*\bar{D}^*}^{(T)}[0]$. In the range $0.8 < \Lambda < 5$ GeV, we cannot find the
bound state solution for $Z^{(T)}_{D^*\bar{D}^*}[1]$ in the OME case.
Thus, we exclude the existence of the
$Z^{(T)}_{D^*\bar{D}^*}[1]$ molecular state.}

\end{enumerate}

\renewcommand{\arraystretch}{1.6}
\begin{table}[hbtp]
\caption{The obtained bound state solutions (binding energy $E$ and root-mean-square radius $r_{\mathrm{RMS}}$) for $D^*\bar{D}^*$ systems.
Here, we discuss two situations, i.e., including all one meson exchange (OME) contribution and only considering one pion exchange (OPE) potential to $D^*\bar{D}^*$ systems. \label{DSDS-1}}
\begin{tabular}{c|cccc|ccc}\toprule[1pt]
&&\multicolumn{3}{c|}{OME}&\multicolumn{3}{c}{OPE}\\\midrule[1pt]
$I^G(J^{PC})$ & State& $\Lambda$ & $E$ (MeV)& $r_{\mathrm{RMS}}$ (fm)& $\Lambda$ & $E$ (MeV)& $r_{\mathrm{RMS}}$ (fm)\\\midrule[1pt]
\multirow{4}*{$1^{+}(0^+)$}&\multirow{4}*{$Z_{D^*D^*}^{(T)}[0]$}
                        &3.6&-0.94&1.74&2.8&-2.03&1.47\\
                        &&3.8&-6.16&1.00&2.9&-6.10&1.00\\
                        &&4&-16.44&0.66&3&-12.51&0.74\\
                        &&4.2&-33.23&0.49&3.1&-21.56&0.59\\\midrule[1pt]
\multirow{4}*{$0^{-}(0^{+-})$}&\multirow{4}*{$Z_{D^*D^*}^{(S)}[0]$}&1.4&-1.72&1.62&3&-5.70&1.24\\
            &&
	    1.5&-17.98&0.88&3.1&-12.15&0.96\\
            &&
	    1.6&-54.60&0.47&3.2&-21.83&0.78\\
\midrule[1pt]
\multirow{4}*{$1^{+}(1^+)$}&\multirow{4}*{$Z_{D^*D^*}^{(T)}[1]$}&\multirow{4}*{-}&\multirow{4}*{-}&\multirow{4}*{-}&4.7&-6.96&0.94\\
                &&           &&&4.8&-12.29&0.73\\
                &&           &&&4.9&-19.36&0.60\\
                &&           &&&5&-28.31&0.51\\\midrule[1pt]
\multirow{4}*{$0^{-}(1^{+-})$}&\multirow{4}*{$Z_{D^*D^*}^{(S)}[1]$}&1.3&-&&3.6&-9.91&1.01\\
                                   &&1.4&-3.44&1.44&3.7&-15.25&0.87\\
                                   &&1.5&-16.57&0.90&3.8&-22.07&0.76\\
                                   &&1.6&-41.25&0.66&3.9&-30.53&0.68\\
\midrule[1pt]
\multirow{1}*{$1^{+}(2^+)$}&\multirow{1}*{$Z_{D^*D^*}^{(T)}[2]$}&-&-&-&-&-&\\
\midrule[1pt]
\multirow{4}*{$0^{-}(2^{+-})$}&\multirow{4}*{$Z_{D^*D^*}^{(S)}[2]$}&1.1&-0.61&1.72&1.6&-3.89&1.28\\
                                   &&1.2&-7.50&1.19&1.7&-9.64&0.98\\
                                   &&1.3&-19.22&0.89&1.8&-18.38&0.77\\
                                   &&1.4&-35.93&0.73&1.9&-30.71&0.64\\
\bottomrule[1pt]

\end{tabular}
\end{table}

\section{Summary}\label{sec4}

Stimulated by the newly observed bottomonium-like states
$Z_b(10610)$ and $Z_b(10650)$, we have carried out a systematical
study of the $B\bar{B}^*$ and $B^*\bar{B}^*$ system using the one
boson exchange model in our work. We have considered both the
S-wave and D-wave interaction between the $B^{(*)}$ and
$\bar{B}^{*}$ mesons, which results in the mixing of the $S$-wave
and $D$-wave contribution as discussed in Sec. \ref{sec2}. Our
numerical results indicate that the $Z_b(10610)$ and $Z_b(10650)$
signals can be interpreted as the $B\bar{B}^*$ and $B^*\bar{B}^*$
molecular states with $I^G(J^P)=1^+(1^+)$ respectively.

As a byproduct, we also predict the existences of six other
$B\bar{B}^*$ and $B^*\bar{B}^*$ bound states (see Table
\ref{coupling}) within the same framework. We want to stress that
the long-range interaction between the heavy meson pair arises
from the one-pion-exchange force, which is clearly known. This OPE
force alone is strong enough to form the above loosely bound
molecular states, which makes the present results quite
model-independent and robust.

The observation of these $Z_b(10610)$ and $Z_b(10650)$ states
shows that the hidden-bottom decay are very important decay
channels, which is characteristic and helpful to the search of the
molecular bottomonium. After taking into account of the phase
space \cite{Nakamura:2010zzi,:2008vj,:2009pz,Adachi:2011ji} and
the conservation of quantum number, the $Z_{B\bar{B}^*}^{(S)}$, ${Z_{B\bar{B}^*}^{(S)}}^\prime$, $Z_{B^*\bar{B}^*}^{(T)}[0]$,
$Z_{B^*\bar{B}^*}^{(S)}[0]$,
$Z_{B^*\bar{B}^*}^{(S)}[1]$ and $Z_{B^*\bar{B}^*}^{(S)}[2]$ molecular states can decay into
\begin{eqnarray*}
&&\big\{\Upsilon(1S)\eta,\Upsilon(2S)\eta,h_b(1P)\eta,\eta_b(1S)\omega\big\},\\
&&\big\{\Upsilon(1S)\omega,\chi_{b0}(1P)\eta,\chi_{b1}(1P)\eta,\chi_{b2}(1P)\eta\big\},\\
&&\big\{\chi_{b1}(1P)\pi,\chi_{b1}(2P)\pi,\Upsilon(1S)\rho,\eta_b(1S)\pi\big\},\\
&&\big\{\Upsilon(1S)\omega,\chi_{b1}(1P)\eta,\eta_b(1S)\eta\big\},\\
&&\big\{\chi_{b0}(1P)\omega, \Upsilon(1S)\eta, \Upsilon(2S)\eta,\eta_b(1S)\omega, h_b(1P)\eta\big\},\\&&\big\{\Upsilon(1S)\omega,\chi_{b1}(1P)\eta,\chi_{b2}(1P)\eta,\eta_b(1S)\eta\big\},
\end{eqnarray*}
respectively. The above modes can be used in the future
experimental search of the partner states of $Z_b(10610)$ and
$Z_b(10650)$.

\renewcommand{\arraystretch}{1.6}
\begin{table*}[htbp]
\caption{A summary of the $B\bar{B}^*$, $B^*\bar{B}^*$,
$D\bar{D}^*$, $D^*\bar{D}^*$ systems. Here, we use $\checkmark$
and $\times$ to mark the corresponding systems with and without
the bound states solution when taking a reasonable $\Lambda$
value, respectively. The criteria of the choice of the reasonable
$\Lambda$ may be strongly biased. \label{coupling}}
\begin{tabular}{c|ccc|ccc}\toprule[1pt]
$I^G(J^P)$&System&Remark&Experiment
\cite{Collaboration:2011gj}&System&Remark&Experiment \\\midrule[1pt]
$1^+(1^+)$&$Z_{B\bar{B}^*}^{(T)}$&$\checkmark$&$Z_b(10610)$&$Z_{D\bar{D}^*}^{(T)}$&$\times$&\\
$0^-(1^{+-})$&$Z_{B\bar{B}^*}^{(S)}$&$\checkmark$&&$Z_{D\bar{D}^*}^{(S)}$&$\checkmark$&\\
$1^-(1^+)$&${Z_{B\bar{B}^*}^{(T)}}^\prime$&$\times$&&${Z_{D\bar{D}^*}^{(T)}}^\prime$&$\times$&\\
$0^+(1^{++})$&${Z_{B\bar{B}^*}^{(S)}}^\prime$&$\checkmark$&&${Z_{D\bar{D}^*}^{(S)}}^\prime$&$\checkmark$&$X(3872)$ \cite{Choi:2003ue}\\
$1^-(0^+)$&$Z_{B^*\bar{B}^*}^{(T)}[0]$&$\checkmark$&&$Z_{D^*\bar{D}^*}^{(T)}[0]$&$\times$&\\
$0^+(0^{++})$&$Z_{B^*\bar{B}^*}^{(S)}[0]$&$\checkmark$&&$Z_{D^*\bar{D}^*}^{(S)}[0]$&$\checkmark$&$Y(3930)$ \cite{:2008vj,:2009pz,Adachi:2011ji}\\
$1^+(1^{+})$&$Z_{B^*\bar{B}^*}^{(T)}[1]$&$\checkmark$&$Z_b(10650)$&$Z_{D^*\bar{D}^*}^{(T)}[1]$&$\times$&\\
$0^-(1^{+-})$&$Z_{B^*\bar{B}^*}^{(S)}[1]$&$\checkmark$&&$Z_{D^*\bar{D}^*}^{(S)}[1]$&$\checkmark$&\\
$1^-(2^+)$&$Z_{B^*\bar{B}^*}^{(T)}[2]$&$\times$&&$Z_{D^*\bar{D}^*}^{(T)}[2]$&$\times$&\\
$0^+(2^{++})$&$Z_{B^*\bar{B}^*}^{(S)}[2]$&$\checkmark$&&$Z_{D^*\bar{D}^*}^{(S)}[2]$&$\checkmark$&$Y(3940)$ \cite{:2008vj,:2009pz,Adachi:2011ji}\\
\bottomrule[1pt]
\end{tabular}
\end{table*}

We also extend our formalism to study the molecular charmonia. The
observed possible molecular charmonia are listed in Table
\ref{coupling}. The possible hidden-charm decay channels of the
molecular states $Z_{D\bar{D}^*}^{(S)}$, $Z_{D^*\bar{D}^*}^{(S)}[0]$,
$Z_{D^*\bar{D}^*}^{(S)}[1]$ and $Z_{D^*\bar{D}^*}^{(S)}[2]$ are
\begin{eqnarray*}
&&\big\{\eta_c(1S)\omega, J/\psi(1S)\eta\big\},\\
&&\big\{J/\psi\omega,\eta_c(1S)\eta\big\},\\
&&\big\{\eta_c(1S)\omega, J/\psi(1S)\eta\big\},\\
&&\big\{J/\psi(1S)\omega, \eta_c(1S)\eta\big\},
\end{eqnarray*}
respectively. Due to the limit of phase space, the hidden-charm
decays for the other one ${Z_{D\bar{D}^*}^{(S)}}^\prime$ molecular state are $J/\psi(1S)$ or
$\eta_c(1S)$ plus multi-pions.

\vfil
\section*{Acknowledgment}

This project is supported by the National Natural Science
Foundation of China (Grants No. 11075004, No. 11021092, No.
11035006, No. 11047606, No. 10805048), and the Ministry of Science
and Technology of China (No. 2009CB825200), and the Ministry of
Education of China (FANEDD under Grants No. 200924, DPFIHE under
Grants No. 20090211120029, NCET under Grants No. NCET-10-0442, the
Fundamental Research Funds for the Central Universities under
Grants No. lzujbky-2010-69).

\end{document}